\documentclass[reprint,amsmath,amssymb,aps,superscriptaddress,pre]{revtex4-2}
\usepackage{graphicx}
\usepackage{color}
\usepackage{hyperref}
\usepackage{longtable}
\usepackage{booktabs}
\usepackage{siunitx}
\usepackage{bm}

\newcommand{\figref}[1]{Figure~\ref{#1}}
\renewcommand{\eqref}[1]{Eq.~(\ref{#1})}
\newcommand{\appref}[1]{Appendix~\ref{#1}}

\newcommand{\kT}{k_{\text{B}}T}

\begin{document}

\title{Predicting Heteropolymer Phase Separation Using Two-Chain Contact Maps}
\author{Jessica Jin}
\affiliation{Department of Chemistry, Princeton University, Princeton, NJ 08544, USA}
\affiliation{Department of Chemical and Biological Engineering, Princeton University, Princeton, NJ 08544, USA}
\author{Wesley Oliver}
\affiliation{Department of Chemical and Biological Engineering, Princeton University, Princeton, NJ 08544, USA}
\author{Michael A. Webb}
\affiliation{Department of Chemical and Biological Engineering, Princeton University, Princeton, NJ 08544, USA}
\author{William M. Jacobs}
\email{wjacobs@princeton.edu}
\affiliation{Department of Chemistry, Princeton University, Princeton, NJ 08544, USA}
\date{\today}

\begin{abstract}
Phase separation in polymer solutions often correlates with single-chain and two-chain properties, such as the single-chain radius of gyration, $R_{\text{g}}$, and the pairwise second virial coefficient, $B_{22}$.
However, recent studies have shown that these metrics can fail to distinguish phase-separating from non-phase-separating heteropolymers, including intrinsically disordered proteins (IDPs).
Here we introduce an approach to predict heteropolymer phase separation from two-chain simulations by analyzing contact maps, which capture how often specific monomers from the two chains are in physical proximity.
Whereas $B_{22}$ summarizes the overall attraction between two chains, contact maps preserve spatial information about their interactions.
To compare these metrics, we train phase-separation classifiers for both a minimal heteropolymer model and a chemically specific, residue-level IDP model.
Remarkably, simple statistical properties of two-chain contact maps predict phase separation with high accuracy, vastly outperforming classifiers based on $R_{\text{g}}$ and $B_{22}$ alone.
Our results thus establish a transferable and computationally efficient method to uncover key driving forces of IDP phase behavior based on their physical interactions in dilute solution.
\end{abstract}

\maketitle

\section{Introduction}
Physics-based metrics that capture the effective interactions among polymers or colloids in solution are commonly used to predict the phase behavior of macromolecular solutions.
For example, the radius of gyration, $R_{\text{g}}$, of a polymer chain can be used to predict the phase behavior of a homopolymer solution because both properties are governed by the same monomer--monomer and monomer--solvent interactions~\cite{colby-rubinstein}.
Similarly, the second virial coefficient, $B_{22}$, which characterizes the effective two-body interactions between polymers or colloidal particles in solution~\cite{mcmillan1945virial, prausnitz1998molecular, vliegenthart2000predicting}, strongly correlates with the phase behavior of these systems, famously leading to a generalized law of corresponding states for colloidal solutions~\cite{vliegenthart2000predicting, noro2000extended}.
However, in the case of chemically complex heteropolymers, such as intrinsically disordered proteins (IDPs), these traditional metrics may fail to accurately predict phase separation due to complex sequence-dependent effects~\cite{martin2020intrinsically, dignon2020biomolecular, borcherds2021intrinsically, sundaravadivelu2024sequence}

Previous studies have shown that order parameters derived from heteropolymer sequences, including simple copolymer sequences~\cite{flory1955theory, pandav2012phase, das2013conformations, sawle2015theoretical, robichaud2019phase, das2018coarse, hazra2020charge, statt2020model, rana2021phase} and more chemically complex IDP sequences~\cite{mao2010netcharge, das2015relating, nott2015phase, lin2016sequence, wang2018molecular, dignon2019temperature, martin2020valence,  nilsson2020finite, perdikari2021predictive}, can distinguish phase-separating from non-phase-separating heteropolymer solutions in situations where $R_{\text{g}}$ and $B_{22}$ are insufficiently discriminating.
Yet such \textit{sequence-based} metrics are typically specific to a particular chemical space or simulation model.
For example, the sequence hydropathy decoration (SHD) parameter~\cite{zheng2020hydropathy}, which describes the patterning of amino-acid hydrophobicity values~\cite{urry1992hydrophobicity, kapcha2014hydrophobicity} in a polypeptide sequence, has been used to classify phase-separating versus non-phase-separating heteropolymers based on their sequences~\cite{an2024active, liao2024sequence, von2024prediction}.
However, relationships between SHD and phase behavior are not readily generalized to different heteropolymer chemistries.
Moreover, because sequence-based metrics do not directly describe physical interactions among the monomers in solution, they provide limited insight into the physical mechanisms that govern heteropolymer phase behavior.
There is thus a pressing need for more discriminating \textit{physics-based} predictors of heteropolymer phase separation, which are sequence-agnostic and can thus be applied across diverse heteropolymer chemistries.
To be of practical use, such predictors should be easily computed based on the interactions between only one or two heteropolymer chains.

We address this challenge with an approach that utilizes statistical features derived from two-chain contact maps, which capture the probability of specific monomer--monomer contacts when two chains are in close proximity. Unlike $B_{22}$, which distills information regarding pairwise heteropolymer interactions into a single number, contact maps retain spatial information regarding the interactions between heteropolymers.
By extracting simple yet informative statistical descriptors from these contact maps, we are able to develop highly accurate phase-behavior classifiers that are both computationally efficient and independent of the chemical details or coarse-graining resolution of a particular heteropolymer model.
Importantly, these classifiers allow us to predict whether a heteropolymer solution with strongly attractive interchain interactions will undergo macroscopic phase separation or instead form finite-size aggregates, a distinction that $B_{22}$ may not capture when sequence-patterning effects are present.

\begin{figure*}
    \centering
    \includegraphics[width=0.85\textwidth]{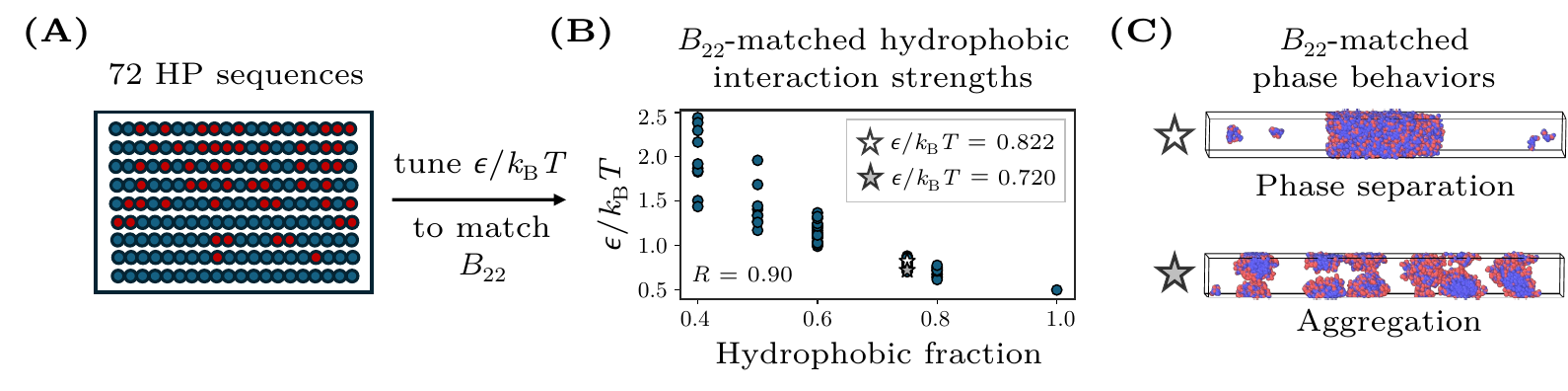}
    \caption{Construction of a $B_{22}$-matched dataset of hydrophobic--polar (HP) copolymers.
      (A) Example HP sequences with varying fractions and arrangements of H and P monomers.
      (B) The hydrophobic interaction strength, $\epsilon/\kT$, is tuned separately for each sequence to achieve a uniform $B_{22}$ value (here, $B_{22} = -1000\sigma^3$). Sequences with lower hydrophobic content generally require stronger $\epsilon/\kT$ (Pearson correlation coefficient $R = 0.90$). The star symbols indicate the sequences corresponding to the simulation snapshots shown in panel C.
      (C) Representative snapshots from direct coexistence simulations illustrating phase separation into coexisting dense and dilute phases (top) versus aggregation into finite-sized clusters (bottom).}
    \label{fig:1}
\end{figure*}

In what follows, we demonstrate the generalizability of our approach by training phase-separation classifiers for two heteropolymer models that differ in their chemical complexity.
First, we consider a hydrophobic--polar (HP) copolymer model and show that contact-map statistics outperform traditional metrics, in particular in scenarios where $B_{22}$ is identical across sequences with divergent phase behaviors.
We also examine how spatial information that is encoded at different length scales in the contact maps contributes to the accuracy of the phase behavior predictions.
We then turn to a dataset of IDP sequences modeled within a more chemically specific coarse-grained framework~\cite{regy2021improved}.
Unlike $R_{\text{g}}$ and $B_{22}$, which are insufficient for classifying the sequences in this dataset, we show that our contact-map approach achieves near-perfect prediction accuracy.
Taken together, our results establish contact-map statistics as robust, generalizable, and computationally efficient predictors of heteropolymer phase separation, providing valuable insights for the design of polymeric materials and biomolecular systems that undergo phase separation.

\section{Results}

\subsection{Hydrophobic--polar copolymers with identical $B_{22}$ values exhibit divergent phase behaviors}

We first investigate the phase behavior of a Kremer--Grest copolymer model~\cite{kremer1990dynamics} in implicit solvent.
In this model, hydrophobic (H) monomers attract one another via a Lennard--Jones interaction with well depth $\epsilon$~\cite{lennardjones1931cohesion}.
By contrast, polar (P) monomers interact with all other H and P monomers via a purely repulsive Weeks--Chandler--Anderson potential~\cite{weeks1971wca}.
All monomers have diameter $\sigma$, and all chains have length $N=20$.
(See \appref{app:models} for a detailed description of the HP model.)

To isolate the effects of sequence heterogeneity on phase behavior, we construct $B_{22}$-matched datasets, in which the dimensionless interaction parameter $\epsilon/\kT$ is tuned independently for each copolymer sequence to achieve a prescribed second virial coefficient, $B_{22}$.
This approach allows us to control for the effective two-body interactions between polymer chains in dilute solution.
In practice, we compute the potential of mean force (PMF), $u(r)$, as a function of the distance, $r$, between the centers of mass of two chains with the same sequence.
We then use the PMF to calculate $B_{22}$,
\begin{equation}
  \label{eq:B22}
  B_{22} = 2\pi \int_{0}^{\infty} \left[1 - \exp\left(-\frac{u(r)}{\kT}\right)\right] r^2 \, dr.
\end{equation}
Negative values of $B_{22}$ indicate net attractive interactions between chains in dilute solution. Given that phase separation requires attractive interactions, we choose sufficiently negative $B_{22}$ values as part of the $B_{22}$-matching procedure.
For every sequence, we adjust $\epsilon/\kT$ to match a target second virial coefficient of either $B_{22}=-400\sigma^3$ or $B_{22}=-1000\sigma^3$ to within statistical error.
We apply this methodology to a set of 75 HP sequences, including sequences used in prior studies~\cite{rana2021phase, statt2020model} as well as additional sequences chosen to balance the $B_{22}$-matched datasets with respect to phase-separating versus non-phase-separating sequences (\figref{fig:1}A and \textit{Supplemental Information}).
In general, sequences with fewer hydrophobic monomers require stronger hydrophobic interactions to match the $B_{22}$ of more hydrophobic sequences.
However, this relationship is not strictly monotonic, as specific sequence patterns can influence the required value of $\epsilon/\kT$ (\figref{fig:1}B).

\begin{figure*}
    \centering
    \includegraphics[width=0.85\textwidth]{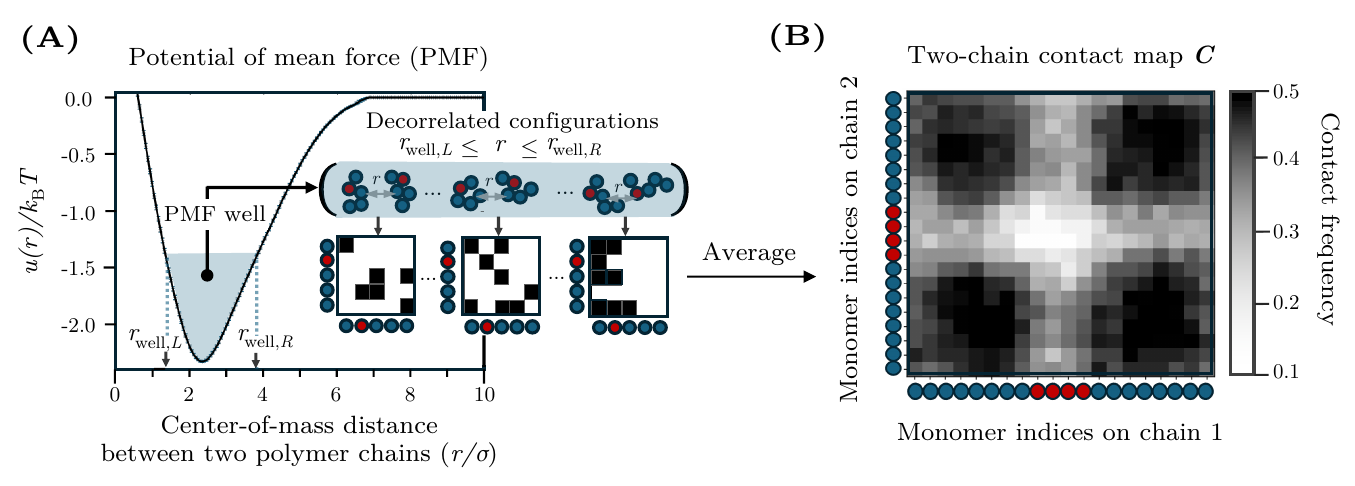}
    \caption{Calculation of two-chain contact maps.
      (A) The potential of mean force (PMF) between two HP polymers as a function of the distance between the centers of mass of the two chains, $r$.
      The PMF, $u(r)$, exhibits an attractive well, defined as the range of distances $r_{\text{well,L}} \le r \le r_{\text{well,R}}$ for which $u(r)$ is within $1\,\kT$ of the minimum.
      Statistically independent configurations from the PMF well are used to construct binary contact matrices in which each element is equal to 1 if monomers $i$ and $j$ are within a specified cutoff distance and 0 otherwise.
      Averaging these contact matrices yields the final contact map, $\bm{C}$.
      (B) An example contact map, in which darker shading indicates higher contact frequencies.
      Schematics along each axis indicate the sequence of H and P residues in the chains.} \label{fig:2}
\end{figure*}

Despite having identical homotypic $B_{22}$ values within each dataset, these heteropolymers exhibit divergent phase behaviors.
Specifically, sequences either phase separate into coexisting dense and dilute phases when the overall polymer concentration is above a saturation concentration or, alternatively, aggregate into finite-sized clusters regardless of the overall polymer concentration (\figref{fig:1}C).
These classifications of sequences as either phase separating or aggregating, which will serve as the ground truth in the analyses that follow, are determined using direct-coexistence simulations in a slab geometry (see \appref{app:classification}).
In the $B_{22} = -400\sigma^3$ dataset, 22 out of 75 sequences (29\%) phase separate, whereas in the $B_{22} = -1000\sigma^3$ dataset, 30 out of 72 sequences whose phase behavior could be determined (42\%) phase separate.
These datasets are thus sufficiently well balanced for the evaluation of the predictive models that we consider next.

The fact that HP sequences with identical $B_{22}$ values can display markedly different phase behaviors highlights a limitation of using $B_{22}$ as the sole predictor of phase separation in heteropolymer solutions, as suggested by prior studies~\cite{statt2020model, das2018coarse, lin2016sequence}.
These differences in phase behavior indicate that factors beyond the average two-body interaction between two isolated chains affect the many-body interactions among chains at high concentrations.

\subsection{Contact-map variance outperforms $B_{22}$ and $R_{\text{g}}$ in predicting phase separation}

To determine whether two-chain simulations contain sufficient information to predict phase separation at higher concentrations, we compute and analyze interchain contact maps.
Because we consider homotypic solutions in which the polymers have the same sequence, the contact maps are symmetric matrices with entries $C_{ij}$, in which each element represents the frequency of contacts between monomer $i$ on one chain and monomer $j$ on another chain.
These contact maps are calculated from the same two-chain simulations with which we compute $B_{22}$.
Specifically, we sample decorrelated configurations for which the center of mass distance, $r$, lies within the attractive well of the PMF (\figref{fig:2}A; see \appref{app:contact-maps}).
Then, for each pair of monomers $(i, j)$, we calculate the fraction of these configurations in which the distance between monomers is less than $r_{\text{c}} = 3\sigma$, which corresponds to the cutoff distance of the Lennard-Jones potential used to model hydrophobic interactions.
Consequently, these contact maps contain information about the spatial distribution of chain--chain contacts.
An example contact map is shown in \figref{fig:2}B, where the axes represent the monomer indices along each polymer chain.
In this contact map, darker regions at the corners indicate preferential interactions between chain ends, whereas lighter interior regions indicate weaker or incidental contacts.

\begin{figure*}
    \centering
    \includegraphics[width=0.75\textwidth]{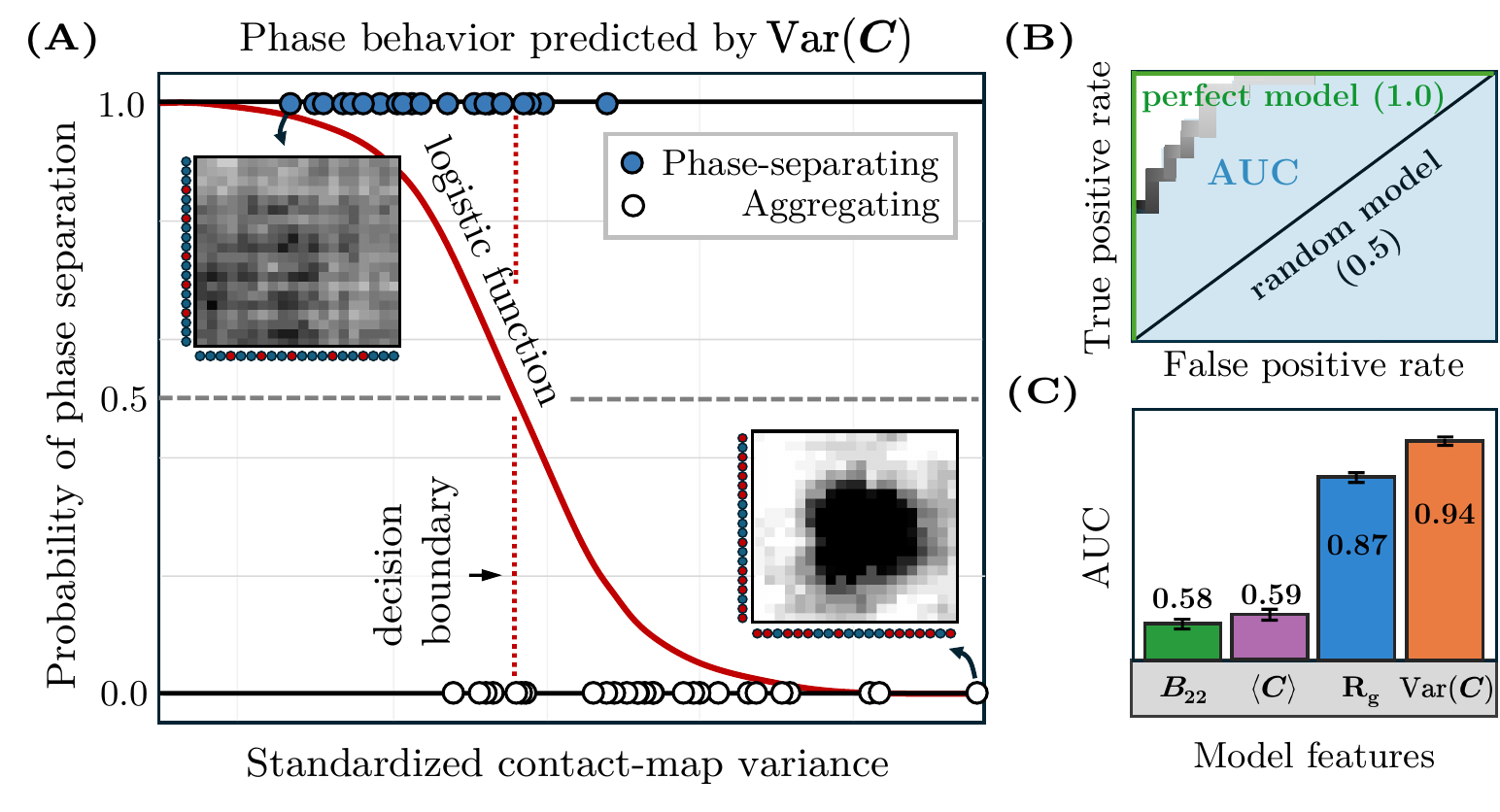}
    \caption{Evaluation of phase-separation predictors via logistic regression.
      (A) A single-variate logistic regression model trained on the contact-map variance for the $B_{22}$-matched dataset with $B_{22} = -1000\, \sigma^3$.
      Each point represents a sequence, plotted according to standardized values of the contact-map variance and colored by its phase behavior.
      The sigmoid function shows how the predicted probability of phase separation varies with contact-map variance.
      The intersection of the sigmoid function and the classification threshold defines the decision boundary, shown as a vertical dashed line, which the model uses to differentiate phase-separating from aggregating sequences. Filled markers (top) that are to the left of this line are correctly classified, whereas those that are to the right are misclassified; the opposite holds for open markers (bottom).
      (B) The receiver operating characteristic curve for the logistic regression model shown in (A) quantifies the trade-off between the true positive and false positive rates as a function of the classification threshold.
      The area under the curve (AUC) quantifies the model's overall performance.
      (C) Performance comparison showing AUC values for $B_{22}$ alone and for bivariate models combining each listed feature with $B_{22}$, using the $B_{22}$-combined dataset.
      Error bars indicate the standard error computed via 185 randomized 80/20 train--test splits.}
    \label{fig:3}
\end{figure*}

Next, we use logistic regression models to evaluate the ability of various metrics to predict phase separation in our $B_{22}$-matched datasets (\figref{fig:3}A).
We select this approach for several reasons. First, logistic regression provides a straightforward and interpretable framework, as it deterministically optimizes a linear decision boundary for classification; this interpretability is advantageous when assessing the relative importance of individual predictors.
Second, given the limited size of our datasets, logistic regression mitigates the risk of overfitting that can arise with more complex, high-parameter models while offering little to no advantage in performance~\cite{harrell2001regression, patel2023data}.
We then evaluate the predictive ability of each logistic regression model by calculating the area under the receiver operating characteristic (AUC), which summarizes the trade-off between the true and false positive rates across all classification thresholds ranging from 0 to 100\% probability of phase separation (\figref{fig:3}B).
An AUC value of 1 indicates perfect prediction, whereas an AUC value of 0.5 is consistent with random guessing.
We determine the uncertainty associated with an AUC value by performing randomized 80/20 train--test splits.
(See \appref{app:logistic-regression} for details of the logistic regression models.)

By comparing the AUCs of simple statistics computed from the contact maps, we identify the contact-map variance, $\text{Var}(\bm{C}) \equiv \sum_{ij} C_{ij}^2 / N^2 - (\sum_{ij} C_{ij} / N^2)^2$, as the most discriminating single-variate predictor of phase separation.
As shown in \figref{fig:3}A, the contact-map variance tends to be higher in aggregating copolymer solutions, whereas more uniform monomer--monomer contact frequencies tend to be predictive of phase separation.
Importantly, this metric significantly outperforms $R_{\text{g}}$ and $B_{22}$, the latter of which is the same across all sequences in a $B_{22}$-matched dataset by construction, and thus has an AUC of 0.5.
The contact-map variance is also more discriminating than the contact-map mean, $\langle \bm{C} \rangle \equiv \sum_{ij} C_{ij} / N^2$.
In the dataset with $B_{22} = -1000\sigma^3$, the contact-map variance achieves an AUC of $0.95 \pm 0.003$ compared to the contact-map mean AUC of $0.72 \pm 0.009$ and the $R_{\text{g}}$ AUC of $0.91 \pm 0.006$.
Similar results are observed in the $B_{22} = -400\sigma^3$ dataset, where the contact-map variance, contact-map mean, and $R_{\text{g}}$ AUC values are $0.91 \pm 0.006$, $0.56 \pm 0.01$, and $0.85 \pm 0.007$, respectively.

We can also generalize this approach to datasets with multiple $B_{22}$ values.
To this end, we combine the two $B_{22}$-matched datasets into a single $B_{22}$-mixed dataset and incorporate $B_{22}$ as an additional feature in a bivariate logistic regression model.
When we use the contact-map variance and $B_{22}$ as the two features, we obtain an AUC of $0.94\pm 0.003$ for the $B_{22}$-mixed dataset.
By comparison, we obtain AUC values of $0.87\pm 0.005$ for the $R_{\text{g}}$-$B_{22}$ bivariate model, $0.59\pm 0.008$ for the contact-map mean-$B_{22}$ model, and $0.58 \pm 0.007$ for the single-variate $B_{22}$ model (\figref{fig:3}C).
The consistency of these results, both for $B_{22}$-matched and $B_{22}$-mixed datasets, suggests that the contact-map variance is a discriminating and robust predictor of phase separation for diverse copolymer sequences across a wide range of hydrophobic interaction strengths.

\subsection{Longest-wavelength contact-map modes are predictive of phase separation}

\begin{figure*}
  \centering \includegraphics[width=0.9\textwidth]{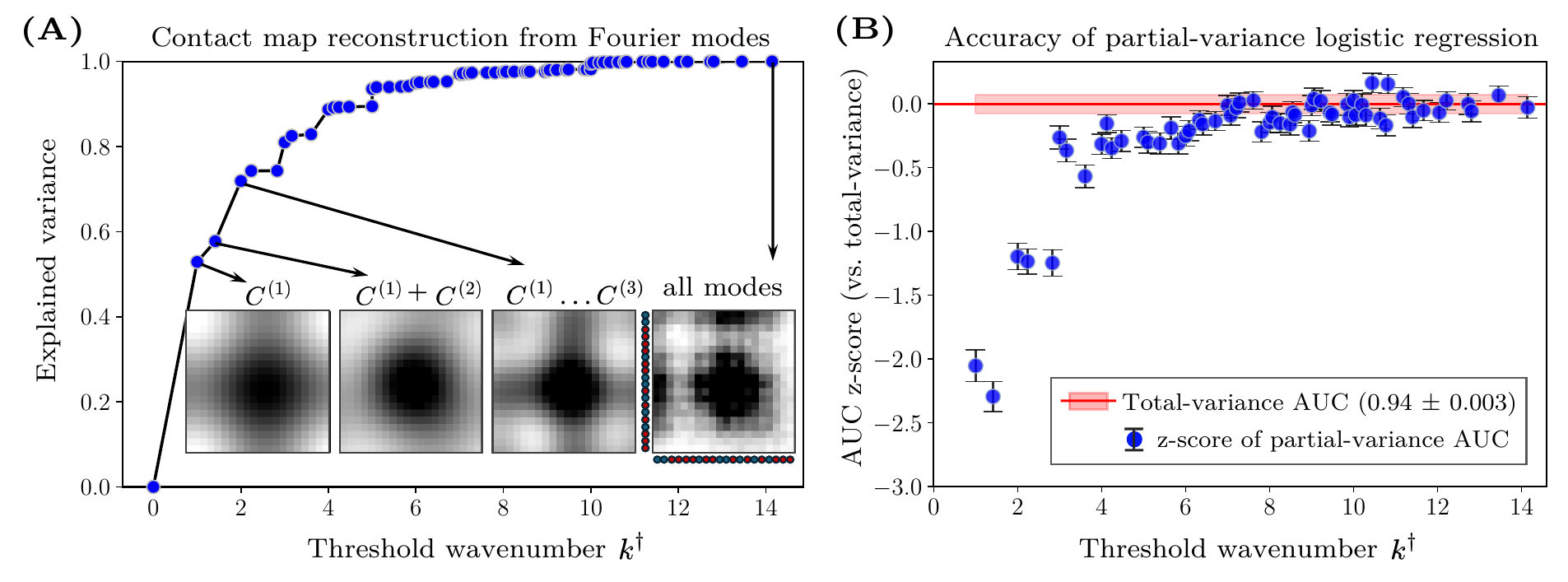} 
\caption{Power spectrum analysis of the contact-map variance and its predictive ability.
  (A) Reconstruction accuracy of an example contact map (sequence ID 2 from the $B_{22}=-400\sigma^3$ dataset) as a function of the threshold wavenumber, $k^\dagger$.
  The explained variance (EV) is defined as $S_{\le k^\dagger} / S$ where $S_{\le k^\dagger}$ is the partial sum of the power values up to wavenumber $k^\dagger$ and $S$ is the total power. Including only the lowest 10\% of the wavenumbers captures approximately 80\% of the total variance.
  Insets show reconstructed contact maps using modes with wavenumbers $k \le k^\dagger$.
  (B) Predictive performance of logistic regression models trained on the $B_{22}$-mixed dataset, which combines the $B_{22} = -400\sigma^3$ and $B_{22} = -1000\sigma^3$ datasets, using $S_{\le k^\dagger}$. To compare with the model trained on $B_{22}$ and the total contact-map variance (AUC = $0.94 \pm 0.003$), we calculate the z-score $[\mathrm{AUC}(S_{\le k^\dagger}) - \mathrm{AUC}(\text{Var})] / \sigma_{\text{Var}}$, where $\sigma_{\text{Var}}$ is the standard deviation of the variance model's AUC scores across train--test splits. Error bars represent standard errors, computed as in \figref{fig:3}C, normalized by $\sigma_{\text{Var}}$.}
\label{fig:4}
\end{figure*}

The contact-map variance can be decomposed into contributions from Fourier modes representing the spatial components of the contact map, where each Fourier mode corresponds to a two-dimensional wavevector, $\vec k = (k_x, k_y)$.
According to Parseval's theorem~\cite{brigham1974fft}, the total contact-map variance is equal to the total power in the Fourier domain,
\begin{equation}
  \text{Var}(\bm{C}) = \frac{1}{N^4} \sum_{k_x,k_y} P(k_x, k_y),
  \label{eq:parseval}
\end{equation}
where $P(k_x, k_y) \equiv |\hat{C}(k_x, k_y)|^2$ is the power associated with wavevector $\vec k$, and the coefficients $\hat{C}(k_x, k_y)$ are the Fourier coefficients of a zero-mean contact map.
Due to the symmetry of the contact map, we consider wavenumbers $k = \sqrt{k_x^2 + k_y^2}$ in practice.
This decomposition allows us to determine the contribution from each wavenumber to the overall contact-map variance.
We can therefore distinguish contributions due to large-scale spatial structure at small $k$ from finer-scale details at large $k$.

By analyzing the power spectra of the contact maps, we find that small-$k$ modes capture most of the total contact-map variance.
We define $S_{\le k^\dagger}$ to be the sum of the power values up to a threshold wavenumber, $k^\dagger$, by restricting the sum over $k$ in \eqref{eq:parseval} to $k \le k^\dagger$.
For most contact maps, $S_{\le k^\dagger}$ approaches the total variance when $k^\dagger$ is small, indicating that most of the power is typically concentrated in the longest-wavelength modes.
Reconstructed contact maps, calculated using an inverse Fourier transform of all modes with $k \le k^\dagger$, illustrate how the spatial structure emerges with increasing $k^\dagger$ (\figref{fig:4}A, inset).
This analysis highlights the dominant role of large-scale spatial structure in determining the total variance of monomer--monomer contact frequencies.

Motivated by this observation, we systematically train logistic regression models using the partial sums, $S_{\le k^\dagger}$, and compare their performance to models trained on the total contact-map variance, $\text{Var}(\bm{C})$.
The predictive ability of these models on the $B_{22}$-mixed dataset quickly approaches that of the total contact-map variance, achieving statistically similar accuracy using only the lowest $10\%$ of the wavenumbers (\figref{fig:4}B).
Beyond $k^\dagger \approx 6$, the inclusion of higher-wavenumber modes offers statistically insignificant improvement, suggesting that, in general, long-wavelength modes contain essentially the same information for predicting phase separation as the total contact-map variance.
These observations indicate that phase separation in these systems is primarily driven by relatively coarse properties of the heteropolymer sequences, as opposed to specific local variations in the monomer--monomer contact frequencies between chains.

\subsection{Accounting for contact-map spatial structure improves discriminatory power}

We next consider whether the spatial structure of a contact map confers additional discriminatory power for predicting phase separation.
Differences in spatial structure are reflected in the power spectra, which can differ among contact maps with identical total variance.
To this end, we extend our analysis by partitioning the power spectrum into two partial sums, the aforementioned $S_{\le k^\dagger}$ and the residual $S_{> k^\dagger}$ that accounts for all modes above $k^\dagger$.
In principle, this ``split-sum'' approach allows us to distinguish between contributions from long-wavelength versus short-wavelength variations when $S_{\le k^\dagger}$ and $S_{> k^\dagger}$ are treated as separate features in a logistic regression model.
Moreover, since $S_{\le k^\dagger}$ and $S_{> k^\dagger}$ sum to the total variance, a split-sum logistic regression model is expected to perform at least as well as the total-variance model because logistic regression operates on linear combinations of the supplied features.

\begin{figure}
    \centering
    \includegraphics[width=\linewidth]{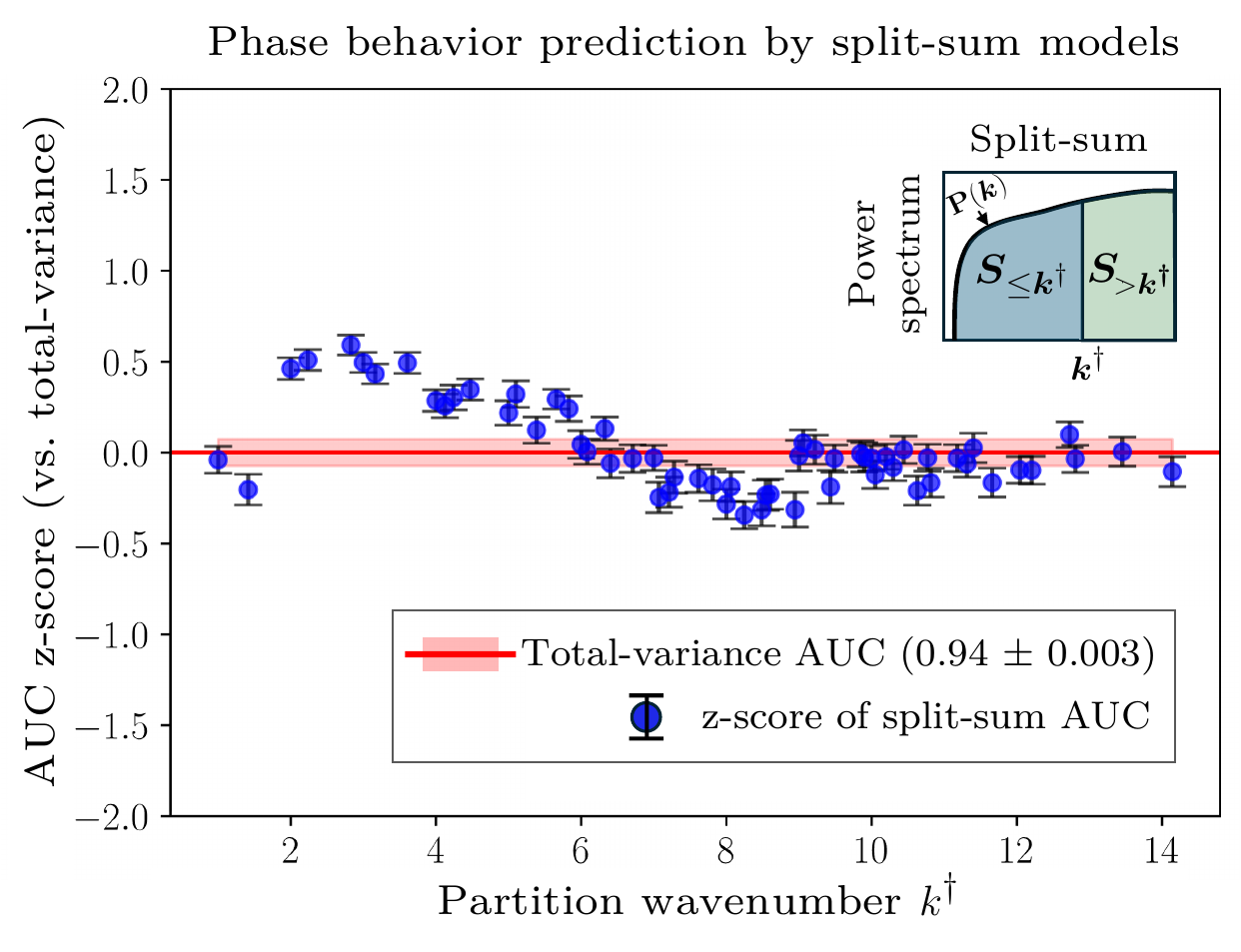}
    \caption{Performance of split-sum logistic regression models (shown schematically in the inset) applied to the $B_{22}$-mixed dataset, showing the AUC scores of logistic regression models trained using $S_{\le k^\dagger}$, $S_{> k^\dagger}$, and $B_{22}$ as features. AUC z-scores and error bars are defined as in \figref{fig:4}B. The split-sum models show statistically significant improvement relative to the total-variance model over a broad range of intermediate $k^\dagger$ values, achieving an average AUC of $0.96 \pm 0.002$ across the range $2 \lesssim k^\dagger \lesssim 4$.}
    \label{fig:5}
\end{figure}

We systematically train split-sum logistic regression models by varying $k^\dagger$ across the power spectrum.
For the $B_{22}$-mixed dataset, we find statistically significant improvement in the AUC using the split-sum approach in an optimal band of $k^\dagger$ values around $2 \lesssim k^\dagger \lesssim 4$ (Figure~\ref{fig:5}).
Specifically, the split-sum models achieve an average AUC value of $0.96 \pm 0.002$ compared to the total-variance AUC value of $0.94 \pm 0.003$.
The consistent improvement of the split-sum models across this relatively broad range of $k^\dagger$ values further indicates that this finding is robust with respect to the partition wavenumber.
However, the details differ slightly among the two $B_{22}$-matched datasets.
For the $B_{22}=-400 \sigma^3$ dataset, splitting the power spectrum yields a notable increase in AUC relative to the total-variance model for $2 \lesssim k^\dagger \lesssim 8$, resulting in an average AUC of $0.96 \pm 0.003$ relative to the total-variance AUC of $0.91\pm 0.006$.
In the case of the $B_{22}=-1000 \sigma^3$ dataset, where the total-variance model already performs exceptionally well ($0.95 \pm 0.003$), a maximum split-sum AUC of $0.97 \pm 0.003$ is reached at $k^\dagger \approx 2$.
Overall, these observations indicate that distinguishing long- and short-wavelength variations provides additional discriminatory power.

We further assess the sensitivity of the split-sum model to the choice of the monomer--monomer contact threshold, $r_{\text{c}}$, for the $B_{22}$-mixed dataset. By sweeping over contact thresholds from $2.0\sigma$ to $4.0\sigma$, we find that the average AUC remains essentially unchanged (0.94--0.96), indicating that our contact-map descriptors are robust with respect to small variations in $r_{\text{c}}$.

The enhanced discriminatory power of the split‑sum approach can be understood by a closer examination of the sequences whose classification changes between the split-sum and total-variance models.
To this end, we analyze the explained variances (EV) of the contact-map Fourier modes.
The wavenumber-ordered EV, $\text{EV}^{(k)}_l$, represents the fraction of the total Fourier power captured by the first $l$ unique modes when ordered by increasing wavenumber.
We contrast the wavenumber-ordered EV with an alternate ordering of the unique wavenumbers in terms of descending power values, $\text{EV}^{(p)}_l$.
When low-$k$ modes dominate, as is most commonly the case, these curves closely overlap.
However, if finer-scale features make unexpectedly large contributions to the total variance, then the EV${}^{(p)}$ curve rises faster than the EV${}^{(k)}$ curve, indicating a more complex power spectrum.
We quantify these differences by measuring the mean-squared difference between EV${}^{(p)}$ and EV${}^{(k)}$, which we use to define the ``variance divergence index'' of a contact map,
\begin{equation}
  \label{eq:vdi}
    \text{Variance divergence index} = \frac{1}{N
    _k} \sum_{l=1}^{N_k} \left(\mathrm{EV}^{(k)}_l - \mathrm{EV}^{(p)}_l\right)^2\!\!,
\end{equation}
where $N_k$ represents the number of unique wavenumbers in the power spectrum.
We then identify sequences in the top 10 percent of this statistic from the $B_{22}$-mixed dataset.
For the sequences in this subset, the split-sum approach significantly outperforms the total-variance model in terms of classification accuracy, defined as the fraction of correct predictions made by the logistic regression model ($0.90 \pm 0.01$ versus $0.76 \pm 0.01$).
By contrast, a control subset of sequences corresponding to the bottom 10 percent of this statistic shows similar classification accuracy between the split-sum and total-variance models ($0.93 \pm 0.0005$). These sequences are provided in the \textit{Supplemental Information}.
This comparison therefore indicates that the split-sum approach is particularly advantageous for classifying sequences whose contact maps are not described by long-wavelength features alone.
This interpretation also explains why the optimal split-sum partition occurs below $k^\dagger \lesssim 6$ (\figref{fig:5}), since these wavenumbers contain most of the information about the total variance of the more common contact maps (\figref{fig:4}).

In summary, our analysis indicates that the variance and spatial distribution of monomer--monomer contacts in two-chain simulations contains additional information---beyond the effective two-body interaction characterized by $B_{22}$---that is predictive of phase separation at higher concentrations.
Although long-wavelength modes typically explain most of the contact-map variance, distinguishing contributions that arise from long- versus short-wavelength variations can further improve the performance of logistic regression models, particularly for datasets containing sequences with unusually large contact-map power values at short wavelengths.

\subsection{Contact-map statistics are more discriminating than $B_{22}$ and $R_{\text{g}}$ in a coarse-grained IDP model}

\begin{figure*}
  \centering \includegraphics[width=0.85\textwidth]{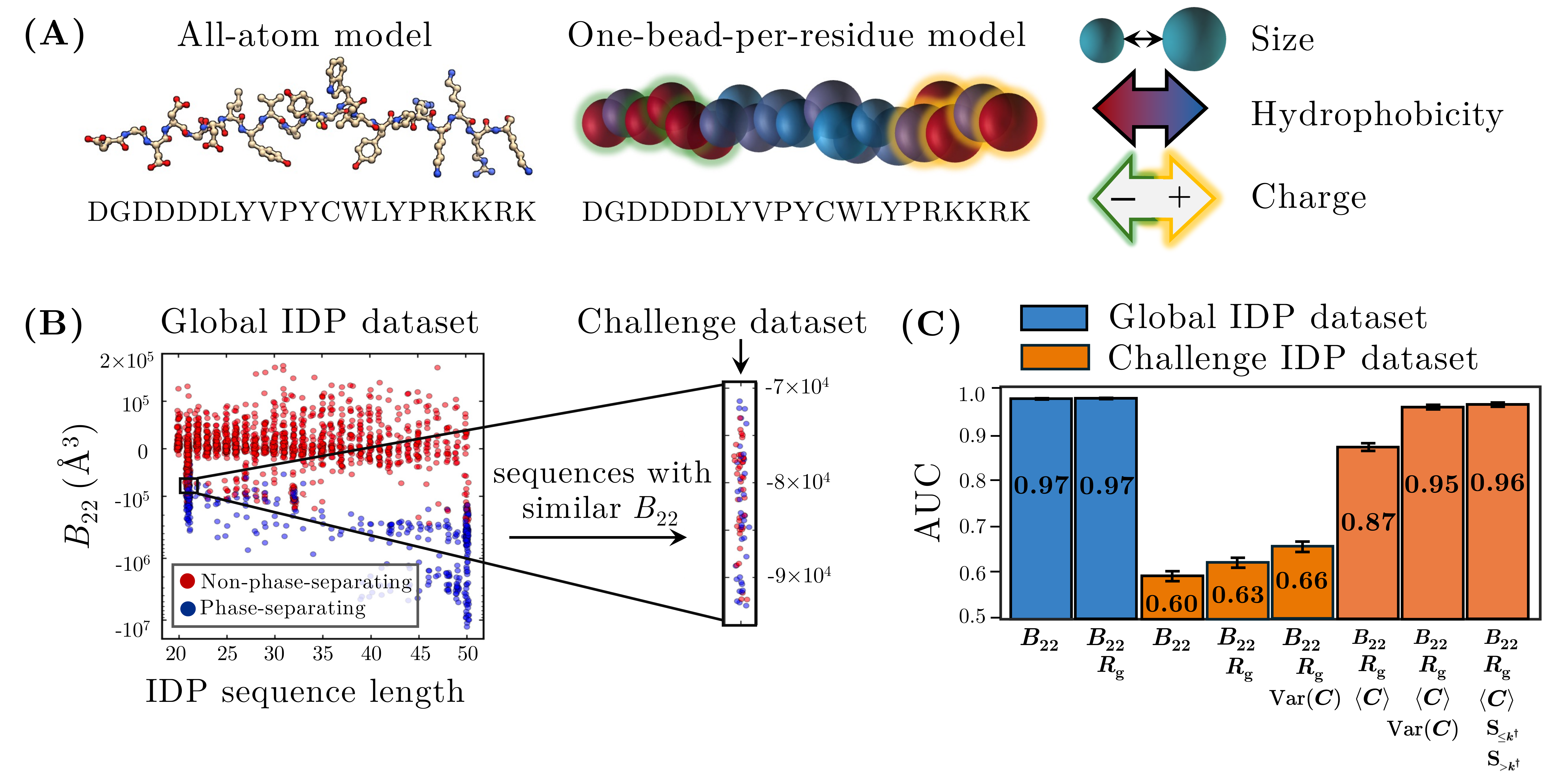}
  \caption{Application of the contact-map approach to a chemically complex heteropolymer model.
    (A) The hydropathy scale (HPS) coarse-grained IDP model~\cite{regy2021improved} represents each residue as a single bead characterized by its size, hydrophobicity, and charge.
    (B) The global IDP dataset spans a range of $B_{22}$ values and sequence lengths.
    The vertical axis uses a logarithmic scale for negative $B_{22}$ values and a linear scale for positive values to capture the full range.
    The highlighted region indicates the ``challenge set,'' a subset of sequences with similar $B_{22}$ values and a fixed length of 21 monomers but differing phase behaviors.
    (C) AUC scores from various logistic regression models applied to both the global IDP dataset and the challenge set, with error bars representing standard errors computed as in \figref{fig:3}C.} 
  \label{fig:6}
\end{figure*}

We now test the transferability of our contact-map analysis to a more complex, chemically specific heteropolymer model.
For this purpose, we consider the hydropathy scale (HPS) coarse-grained (CG) model of IDPs developed by Mittal and colleagues~\cite{regy2021improved}.
This model represents each amino acid residue as a single bead characterized by its size, hydrophobicity, and charge (\figref{fig:6}A).
These properties allow the model to capture key aspects of IDP physical chemistry, including hydrophobic and screened electrostatic interactions (see \appref{app:models}).
The HPS model poses a challenging test for the generalizability of our contact-map approach not only due to the increased complexity of having 20 distinct monomer types but also due to the incorporation of multiple kinds of attractive and repulsive interactions that act over different length scales.

We first analyze a large dataset comprising a diverse set of 2034 IDPs, including both naturally occurring sequences and sequences designed to exhibit a broad range of thermodynamic and dynamic behaviors~\cite{an2024active, yaxin-wes-data-set}.
IDPs in this dataset span a range of sequence lengths from 20 to 50 residues and have $B_{22}$ values ranging from \SI{-10e7}{\AA^3} to \SI{1e5}{\AA^3} (\figref{fig:6}B). A logistic regression model using $B_{22}$ alone achieves excellent predictive performance, with an average AUC of $0.97\pm 0.001$ on unseen test data (\figref{fig:6}C). Including $R_{\text{g}}$ as a additional feature in a bivariate model maintains the same predictive performance ($0.97\pm 0.001$).
This high \textit{overall} accuracy reflects the global trend that IDP sequences with more negative $B_{22}$ values tend to have a greater propensity for phase separation.
This overall behavior is consistent with the results of previous studies using this and similar CG IDP models~\cite{statt2020model, rana2021phase, rekhi2023role, panagiotopoulos2024sequence}.

However, these global trends break down when we consider sequences with highly similar $B_{22}$ values.
To isolate this effect, we examine a curated ``challenge set'' of 75 IDP sequences, each comprising $N=21$ amino acids.
These IDP sequences have negative $B_{22}$ values clustered within a narrow range of \SI{-1e5}{\AA^3} to \SI{-7e4}{\AA^3}.
Despite having similar $B_{22}$ values and identical chain lengths, only 37 of these 75 sequences were previously determined to phase separate~\cite{yaxin-wes-data-set}.
As a result, the discriminatory power of $B_{22}$ and $R_{\text{g}}$ are substantially diminished in the context of this challenge set, with the AUC of a bivariate $B_{22}$ and $R_{\text{g}}$ model dropping to $0.63\pm 0.01$.
Thus, the shortcomings of traditional physics-based metrics that we observed in the simpler copolymer model carry over to this CG IDP model, particularly in situations where sequence-dependent effects appear to play an important role.

As in the simpler copolymer model, incorporating contact-map statistics greatly improves the accuracy of logistic regression models trained on the IDP challenge set.
Following the approach established for the copolymer model, we construct contact maps using configurations from the two-chain PMF well.
For the IDP model, the monomer--monomer contact threshold $r_{\text{c}}$ is treated as a hyperparameter.
We select an optimal value of \SI{24}{\AA} based on the performance of the most accurate logistic regression models, although the classification AUC changes minimally across a broad range of contact thresholds (see \appref{app:contact-maps}).
In this setting, a bivariate model using $B_{22}$ and the contact-map variance as features does not substantially improve the AUC relative to the bivariate $B_{22}$--$R_{\text{g}}$ model.
However, when we introduce the contact-map mean as an additional feature, we observe substantial improvement: a bivariate model using $B_{22}$ and the contact-map mean as features attains an AUC of $0.87\pm 0.01$, whereas a four-feature model utilizing $B_{22}$, $R_{\text{g}}$, and both the contact-map mean and variance improves the AUC to $0.95\pm 0.004$ (\figref{fig:6}C).
For this challenge set, constructing a five-feature model by replacing the total contact-map variance with the two split-sum features further increases the AUC to $0.96\pm0.004$, representing a small yet statistically significant improvement over the four-feature model.

These findings suggest that in more complex heteropolymer models, such as the HPS IDP model, both the mean and the variance of the contact map provide important information for predicting phase separation.
We hypothesize that the simpler behavior of the copolymer model, for which the contact-map mean is less informative, is a consequence of its much simpler pair potential, which has a single length scale for all attractive interactions.
Thus, in general, we anticipate that multiple contact-map statistics will be needed to discriminate phase-separating from non-phase-separating heteropolymers in more chemically complex models.

\section{Discussion}

In this study, we demonstrate that incorporating statistical features derived from two-chain contact maps significantly enhances the prediction of heteropolymer phase separation in systems ranging from simple copolymer models to chemically complex IDP models.
These contact maps capture the physical arrangements of the monomers when the polymer chains are in contact under dilute conditions, retaining crucial information that can be used to distinguish heteropolymer phase behavior in bulk solutions.
Our approach thus provides superior phase-behavior predictions compared to traditional physics-based single-chain and two-chain metrics, while also offering insights into the spatial dependence of the physical interactions driving phase separation.
Most importantly, our approach is transferable across different heteropolymer models because it relies on physical observables as opposed to model-specific sequence features.

The strong relationship between contact-map variance and copolymer phase separation is consistent with the hypothesis that uniform monomer--monomer interactions promote \textit{collective} interactions that lead to phase separation~\cite{martin2020valence}.
By contrast, more heterogeneous interactions---reflected by a greater contact-map variance---tend to create localized ``hot spots'' that stabilize finite-sized clusters but do not promote bulk condensation.
This interpretation is further supported by our finding that long-wavelength modes are typically sufficient to reproduce the discriminatory power of the total contact-map variance, although distinguishing larger-scale interactions from finer-scale variations can improve the results for less common heteropolymer sequences with short-wavelength contact-map features.
Moreover, our results demonstrate that contact-map statistics and power-spectrum analyses provide quantitative insights in situations where there is no obvious way to classify monomer types as being either strongly interacting ``stickers'' or relatively inert ``spacers''~\cite{pappu2023phase}, for example due to a mixture of length scales over which monomers can interact.
Extensions of our contact-map approach may thus prove useful for quantifying sequence heterogeneity more generally, without requiring system-specific classifications of monomer types~\cite{zeng2023developments, martin2020valence}.

Looking forward, our contact-map approach may serve as a useful tool in emerging areas of polymer science.
One exciting direction is the integration of contact-map features into active learning frameworks for polymer design and discovery~\cite{jablonka2021bias, ramesh2023polymer, an2024active, von2024prediction, chang2025mechanical}.
By providing a unifying representation that utilizes physical observables as opposed to sequence features and is thus transferable across models of varying complexity, contact-map statistics could enhance machine learning methods that incorporate training data from models with different fidelities, ranging from atomistic to coarse-grained and field-theoretic simulations.
Moreover, when scaling to more complex models, incorporating additional features does not necessarily reduce the model interpretability, since feature-selection techniques such as L1 regularization can be used to identify the most statistically relevant features~\cite{tibshirani1996regression}.
Our contact map approach could also be extended to multicomponent heteropolymer mixtures, for which the prediction of miscibility and phase separation is an even more demanding challenge~\cite{jacobs2023theory, chen2024emergence}.
However, considering the increased complexity of multicomponent phase diagrams, it will likely be more meaningful to predict the compositions of coexisting phases as opposed to simply distinguishing phase-separated from aggregated states.
Given an appropriate classification objective, our framework could be applied by incorporating statistics from both homotypic and heterotypic interchain contact maps into a unified predictive model.
We note that computing these statistics entails no significant additional computational cost compared to calculating the pairwise $B_{22}$ coefficients for a multicomponent mixture.
Our approach may therefore contribute to the development of computationally efficient methods to design heteropolymer blends with tailored properties and to unravel the sequence determinants of protein and nucleic acid phase separation in living cells.

\begin{acknowledgments}
  We thank Fan Chen for contributing code for the HP model direct-coexistence simulations and for helpful discussions.
  We also thank Yury Polyachenko for insightful comments and Milos Nikolic for feedback on the manuscript draft.
  Research reported in this publication was supported by the National Institute of General Medical Sciences of the National Institutes of Health under award number R35GM155017 to WMJ.
  Simulation and analysis scripts can be obtained at \texttt{https://github.com/wmjac/} \texttt{heteropolymer-phase-separation}.
\end{acknowledgments}

\appendix

\section{Heteropolymer simulation models}
\label{app:models}

\textit{HP copolymer model.} We study HP copolymers using a Kremer--Grest model~\cite{kremer1990dynamics} in implicit solvent.
In our implementation, hydrophobic (H) monomers with diameter $\sigma$ attract one another via a cut-and-shifted Lennard-Jones (LJ) potential~\cite{allen2017computer} with well depth $\epsilon$ and cutoff distance of $3\sigma$.
Polar (P) monomers interact with both H and P monomers via a purely repulsive Weeks--Chandler--Anderson (WCA) potential~\cite{weeks1971role}.
Each polymer chain consists of $N=20$ monomers connected by finite extensible nonlinear elastic (FENE) bonds~\cite{kremer1990dynamics}.

For each sequence, we tune $\epsilon/\kT$ to match the target $B_{22}$ value at a fixed simulation temperature $\kT = 1$.
Specifically, we calculate the potential of mean force (PMF), $u(r)$, for each polymer as a function of the center-of-mass distance $r$ between two identical chains using adaptive biasing force (ABF)~\cite{darve2001calculating} simulations implemented in the COLVARS~\cite{fiorin2013using} module in LAMMPS~\cite{plimpton1995fast}.
We then obtain $B_{22}$ from the PMF using \eqref{eq:B22} and compute an average over four independent trials.
We systematically vary $\epsilon/\kT$ until we match either $B_{22} = -400\sigma^3$ or $B_{22} = -1000\sigma^3$ to within the statistical error of the four trials for each sequence.
All sequences are provided in the \textit{Supplemental Information}.

\textit{HPS IDP model.}
We employ the HPS-Urry coarse-grained (CG) model for intrinsically disordered proteins (IDPs)~\cite{regy2021improved}, in which each amino acid is represented by a single bead characterized by its size, hydrophobicity, and charge.
Hydrophobic interactions are incorporated using hydropathy values derived from the Urry scale~\cite{urry1992hydrophobicity}, whereas electrostatic interactions between charged residues are modeled using Debye-H\"{u}ckel potentials with inverse screening length $\kappa = 1\:\text{nm}^{-1}$ to reflect physiological conditions.
All HPS IDP simulations are conducted at a temperature of \SI{300}{K}.
Sequences are sourced from Ref.~\cite{yaxin-wes-data-set}, which includes a diverse collection of 2034 heteropolymer sequences varying in length from 20 to 50 residues.
We focus on a ``challenge set'' of 75 21-residue sequences, which have similar $B_{22}$ values ranging from approximately \SI{-1.0e5}{\AA^3} to \SI{-7e4}{\AA^3}.
The challenge set sequences are nearly even split between aggregating (38) and phase-separating (37) sequences, as determined by the equation-of-state (EOS) method presented in Ref.~\cite{an2024active}.
We compute homotypic $B_{22}$ values for all IDP sequences in the global dataset following the same approach as for the HP copolymers.
All challenge set sequences are provided in the \textit{Supplemental Information}.

\section{Phase behavior classification}
\label{app:classification}

\textit{Phase behavior classification for HP copolymers.}
We employ direct-coexistence molecular dynamics (MD) simulations~\cite{ladd1977triple} using a slab geometry with periodic boundary conditions at constant temperature $\kT = 1$.
We initialize a $20\sigma \times 20\sigma \times 36\sigma$ pre-equilibrated polymer melt with monomer number density $0.6/\sigma^3$ adjacent to a $20\sigma \times 20\sigma \times 144\sigma$ empty region and allow the system to come to equilibrium. A total of $n=432$ chains are thus present in a simulation box with volume $V=20\sigma \times 20\sigma \times 180\sigma$.
We then analyze the equilibrium density profile along the long dimension of the simulation box to assess whether a stable interface is formed, which is indicative of phase separation.
If phase separation is not observed, we systematically reduce the average number density by increasing the long dimension of the simulation box in increments of $72\sigma$.
This procedure allows us to look for condensed phases with number densities considerably lower than that of a polymer melt.
We terminate this procedure when $n/V$ is less than the overlap number density~\cite{colby-rubinstein}, at which points the chains do not interact strongly in a single phase.
In this case, we classify the sequence as non-phase-separating.

\textit{Phase behavior classification for IDP heteropolymers.}
We classify the IDP phase behavior using the EOS method introduced in Ref.~\cite{an2024active}.
Specifically, we perform constant-volume MD simulations of 100 chains at \SI{300}{K} in a cubic simulation box with fixed volume and periodic boundary conditions.
We then measure the equilibrium pressure for number densities ranging from the overlap number density to the number density of a polymer melt.
Sequences are classified as phase-separating if the EOS exhibits a nonmonotonic ``van der Waals loop''~\cite{binder2012beyond, mcquarrie1976statistical} and non-phase-separating otherwise.

\section{Contact-map construction and analysis}
\label{app:contact-maps}

\textit{Contact map construction and statistics.}
We construct ensemble-averaged contact maps using a consistent approach for both the HP and IDP models.
For each sequence, we compute the PMF for two identical chains as described in \appref{app:models}.
From the PMF, $u(r)$, we define an attractive well based on the range of center-of-mass distances $r$ where $u(r)$ lies within $1\,\kT$ of its minimum value.
For the IDP model, we use the range where $u(r)$ is within $0.4\,\text{kJ/mol}$ of the minimum. This value, about half the average well depth across PMFs, balances isolating configurations where chains strongly interact while maintaining sufficient statistics.
We then sample at least 2400 decorrelated configurations of the two interacting chains from distances within the PMF well to ensure that the resulting ensemble-averaged contact maps are converged.
A contact is defined if the distance between monomers $i$ and $j$ is less than a specified contact threshold $r_{\text{c}}$.
For the HP copolymer model, $r_{\text{c}} = 3\sigma$, which is set equal to the cutoff distance for the nonbonded LJ interactions.
For the IDP model, we treat the contact threshold $r_{\text{c}}$ as a hyperparameter and find an optimal value of \SI{24}{\AA} based on the performance of the four-feature logistic regression model described in the main text.
Notably, the AUC scores of this model are robust with respect to variations in the contact threshold near the optimal value ($0.90 \pm 0.006$ at $r_{\text{c}} = \SI{20}{\AA}$ and $0.94 \pm 0.004$ at $r_{\text{c}} = \SI{28}{\AA}$).

\textit{Fourier analysis.}
To analyze spatial patterns within contact maps, we compute the two-dimensional discrete Fourier transform (DFT) of the zero-meaned, symmetrized contact map, $(C_{ij} + C_{ji})/2 - N^{-2} \sum_{i,j} C_{ij}$, for each polymer sequence.
This operation decomposes the contact map into a set of coefficients $\bm{\hat{C}}(k_x, k_y)$, where $k_x$ and $k_y$ are the components of a wavevector.
Here, the modes indexed by $k_x$ and $k_y$ are standing waves oriented parallel to the horizontal and vertical axes of the contact map, respectively, with each mode capturing a characteristic length scale of correlated monomer--monomer contacts.
  The magnitudes of the corresponding Fourier coefficients reflect how strongly spatial patterns at each length scale are expressed in the contact map.
For real-valued symmetric contact maps, the Fourier transform exhibits conjugate symmetry, $\bm{\hat{C}}(-k_x, -k_y) = \bm{\hat{C}}(k_x, k_y)^*$, as well as diagonal symmetry, $\bm{\hat{C}}(k_x, k_y) = \bm{\hat{C}}(k_y, k_x)$, reducing the number of unique components.

\section{Logistic regression models}
\label{app:logistic-regression}

We use logistic regression to assess the predictive ability of various combinations of physical feature variables, which can be written as an $m$-dimensional feature vector $\mathbf{X} = [x_1, x_2, \dots, x_m]$.
Each sequence in a dataset is classified by a binary variable $y \in \{0, 1\}$, where $y = 1$ corresponds to phase separation and $y = 0$ corresponds to aggregation.
The logistic model relates the log-odds of the conditional probability $P(y \mid \mathbf{X})$ to a linear combination of the features,
\begin{equation}
\log \left( \frac{P(y = 1 \mid \mathbf{X})}{1 - P(y = 1 \mid \mathbf{X})} \right) = \beta_0 + \beta_1 x_1 + \cdots + \beta_m x_m,
\end{equation}
where $\beta_0$ is the bias and the coefficients $\beta_1, \ldots, \beta_m$ weight each feature.
We determine the optimal coefficients $\bm{\beta}$ by maximizing the associated log-likelihood function without additional regularization terms.
Each logistic regression model is trained using a randomized 80/20 train--test split of the dataset.
The model is then evaluated on the test set to compute the area under the receiver operating characteristic (AUC) metric, which is obtained by plotting the true positive rate against the false positive rate as the bias, $\beta_0$, is varied from $-\infty$ to $\infty$.
This approach ensures robustness to slight class imbalances.
We compute an average AUC score and an associated standard error by evaluating 185 randomized train--test splits.

For split-sum models, the total variance model serves as a baseline.
Because the total variance aggregates the same Fourier components used in the split-sum models, no combination of partial sums should, in principle, perform worse.
  Accordingly, the split-sum models are expected to perform at least as well as this baseline, although slight deviations can occur in practice due to overfitting, which can be seen in cases where a model achieves a higher training AUC than the total variance model but fails to generalize to unseen data.
  For this reason, we apply regularization during model selection to penalize marginal training gains, and thus to discourage overfitting,
\begin{equation}
  \text{AUC}_{\text{adj}} = \text{AUC}_{\text{train}} - \frac{\lambda}{|\text{AUC}_{\text{train}} - \text{AUC}_{\text{baseline}}|},
\end{equation}
where $\lambda=6.0\times 10^{-5}$.
This adjustment to the AUC during training is applied to the $B_{22}$--combined dataset, where overfitting is more pronounced than in $B_{22}$--matched datasets.
For the IDP dataset, split-sum models perform comparably without regularization, so unregularized models are used for the reported results.
As noted above, we further evaluate each model over repeated train-test splits to ensure that the observed improvements are generalizable and not a result of overfitting.

\clearpage
\section*{Supplemental Information}

\subsection*{$B_{22}$-matched HP model datasets}
The following tables provide the sequence information and $\epsilon$ values for HP copolymers at constant $B_{22}$ values of $-400\sigma^3$ and $-1000\sigma^3$.
The epsilon values were tuned to achieve the same homotypic $B_{22}$ values (within statistical error) across four trials.
Note that the sequences differ slightly between the tables, because sequences that did not exhibit clear phase separation or aggregation in our direct-coexistence simulations were excluded.
The fraction of H monomers in each sequence is denoted by $x_{\text{H}}$.
PS reports whether the sequence phase separates ($\checkmark = $ yes, X $=$ no).
The complete dataset can be obtained at \texttt{https://github.com/wmjac/} \texttt{heteropolymer-phase-separation}

\begin{longtable}{r|c|c|c|c}
\caption{Sequence information for $B_{22}$ = -400 $\sigma^3$} \\
\toprule
\textbf{Index} & \textbf{Sequence} & \textbf{$x_{\text{H}}$} & \textbf{$\epsilon$} & \textbf{PS} \\
\midrule
\endfirsthead

\toprule
\textbf{Index} & \textbf{Sequence} & \textbf{$x_{\text{H}}$} & \textbf{$\epsilon$} & \textbf{PS} \\
\midrule
\endhead

\midrule
\multicolumn{5}{r}{\textit{Continued on next page}} \\
\midrule
\endfoot

\bottomrule
\endlastfoot

1 & PHPHPPHPPHPPHPPHPHPH & 0.4 & 2.285 & X \\
2 & HHPPPPHPPHHPHPHPHPPP & 0.4 & 1.762 & X \\
3 & HPPHPPHPPHPPHPPHPHPH & 0.4 & 2.222 & X \\
4 & PHPHPPHPPHPPHPPHPPHH & 0.4 & 2.030 & X \\
5 & HPHPHPPHPHPPPPHPHPPH & 0.4 & 2.137 & X \\
6 & PHPPPPPHHHHPHHPPPHPP & 0.4 & 1.385 & X \\
7 & HPPPHHHPHPPHPPPHPPPH & 0.4 & 1.700 & X \\
8 & PPHHHPHPHPPPPPHPHPPH & 0.4 & 1.705 & X \\
9 & PPHPPPHPPHPHHHHHPPPP & 0.4 & 1.317 & X \\
10 & PHHHPPHPPPHPHPPPPHPH & 0.4 & 1.790 & X \\
11 & HPHPHPHPHPHPHPHPHPHP & 0.5 & 1.770 & X \\
12 & HPPHPPHHPHPHHPHPHPHP & 0.5 & 1.550 & X \\
13 & HPHHPPPHPPHHHPPPHPHH & 0.5 & 1.344 & X \\
14 & PPPHHPHHHPHPPHPPHPHH & 0.5 & 1.295 & X \\
15 & PPHHPPHPPHPHHPHPPHHH & 0.5 & 1.320 & X \\
16 & PHPPPHHHPPHHHPHHHPPP & 0.5 & 1.172 & X \\
17 & PPHHHPPHPPHPPHPPHHHH & 0.5 & 1.216 & X \\
18 & HHHHPPHHPPPHPPHHPPHP & 0.5 & 1.216 & X \\
19 & HHHHPPPHHHHPPPPPHPPH & 0.5 & 1.068 & X \\
20 & HPHHPHHHHPHPHPPPPPHP & 0.5 & 1.147 & X \\
21 & HHHPPPHHHPPHHHPPPHHH & 0.6 & 0.970 & \checkmark \\
22 & HPHPPPHHHHPHHHPHPHHP & 0.6 & 1.019 & X \\
23 & PHPHPPHHPHPHHHHPPHHH & 0.6 & 1.000 & X \\
24 & HHHHPPHHPHHHPHPPHPHP & 0.6 & 0.988 & X \\
25 & HHHHPPPHHPHHPHHHHPPP & 0.6 & 0.938 & X \\
26 & HHHHPPHPHHHPPHHHHPPP & 0.6 & 0.937 & X \\
27 & HHHHPPHHHPPHHHHPPPPH & 0.6 & 0.918 & X \\
28 & PHHPHPHHHHHHHPPPPHPH & 0.6 & 0.910 & X \\
29 & HHHPPHHHHHPHHPPHPPHP & 0.6 & 0.926 & X \\
30 & HPPPPHHPHHHHHHHPPHPH & 0.6 & 0.884 & X \\
31 & PHPPPPHHHHHPHHPHPHHH & 0.6 & 0.913 & X \\
32 & HHHHHHPPHHHPPHPPHPHP & 0.6 & 0.886 & X \\
33 & HHHHPHHPHHPHPHHPHHHH & 0.75 & 0.748 & \checkmark \\
34 & HPHPHHHHHHHHHPHPHHHP & 0.75 & 0.699 & X \\
35 & HHPHHPHHHHHHHHHPHHPP & 0.75 & 0.672 & X \\
36 & PHPHHHHHHHHHHHHPHPPH & 0.75 & 0.641 & X \\
37 & HHHHHHHHHHPHHHPHPHPP & 0.75 & 0.638 & X \\
38 & PPHPHPHHPHHHHHHHHHHH & 0.75 & 0.630 & X \\
39 & PPHPPHHHHPHHHHHHHHHH & 0.75 & 0.622 & X \\
40 & PPHPHPHPHHHHHHHHHHHH & 0.75 & 0.614 & X \\
41 & HHHHHHHHPPPPHHHHHHHH & 0.8 & 0.580 & \checkmark \\
42 & PPHHPHHHHHHHHHHHHHPH & 0.8 & 0.586 & X \\
43 & PPHHHHHHHHHHHHHHHHPP & 0.8 & 0.564 & X \\
44 & PPPHHPHHHHHHHHHHHHHH & 0.8 & 0.555 & X \\
45 & PPPPHHHHHHHHHHHHHHHH & 0.8 & 0.536 & X \\
46 & HHHHHHHHHHHHHHHHHHHH & 1.0 & 0.420 & \checkmark \\
47 & PHHHPHPHPHPPPHPPPHHH & 0.5 & 1.300 & X \\
48 & PHPPPHHHHHPHPHPHHPHH & 0.6 & 0.985 & X \\
49 & HHPHHHPHHHPPHPPHPHPH & 0.6 & 1.020 & X \\
50 & HHPHHHHPHPPHHPHPPPHH & 0.6 & 1.000 & X \\
51 & HPHHPHHHPPPHHPHPHPHH & 0.6 & 1.110 & \checkmark \\
52 & HHHHPPPHPHHPHHPHPHPH & 0.6 & 1.053 & X \\
53 & HHPHPHHHPHPHPPPHHPHH & 0.6 & 1.070 & \checkmark \\
54 & HHHHHPPHPPPHHHPPPHHH & 0.6 & 0.915 & X \\
55 & PHHHHPHHPHPHHPPHHPPH & 0.6 & 1.070 & X \\
56 & PHPHHHHPHHPHPPPHHHHP & 0.6 & 1.010 & X \\
57 & HPHHPPHPHPPHHHPPHHHH & 0.6 & 1.010 & X \\
58 & HHPHHPHPPHPHHHPPHPHH & 0.6 & 1.080 & \checkmark \\
59 & HPHHHPHPHPHPPHHPHHHP & 0.6 & 1.120 & X \\
60 & HPHHPHHPPHHPPHHPHHPH & 0.6 & 1.150 & \checkmark \\
61 & HPHHPHPHPHHPHPHPHHPH & 0.6 & 1.200 & \checkmark \\
62 & HPHHPHHPHPHPHHPHHPHP & 0.6 & 1.180 & X \\
63 & HHHPHHPHHHPHHPHHPHHH & 0.75 & 0.766 & \checkmark \\
64 & PHHHPHHHPHHHPHHHPHHH & 0.75 & 0.750 & \checkmark \\
65 & HHHPHPHHHPPHHPHHHHHH & 0.75 & 0.720 & \checkmark \\
66 & HHPHHHHHHPHPHHPHHHHP & 0.75 & 0.735 & \checkmark \\
67 & HHHHHPHPHHPHHHPHHHPH & 0.75 & 0.738 & \checkmark \\
68 & HHHPHPHHHHHHPHHHPHHP & 0.75 & 0.720 & \checkmark \\
69 & HPHHHPHPHHHPHHHPHHHH & 0.75 & 0.750 & \checkmark \\
70 & HHPHPHHHHPHHHHHHHHHP & 0.8 & 0.630 & \checkmark \\
71 & HHPHHHHHHPHHHHHPHHPH & 0.8 & 0.650 & \checkmark \\
72 & HHHHHPHHHPHHHHPHPHHH & 0.8 & 0.650 & \checkmark \\
73 & HPHHPPHHHHHHHHHHPHHH & 0.8 & 0.620 & \checkmark \\
74 & HPPHHHHHPHHHHHHHPHHH & 0.8 & 0.628 & \checkmark \\
75 & HHHPHHHPHHHHPHHHPHHH & 0.8 & 0.670 & \checkmark \\
\end{longtable}

\begin{longtable}{r|c|c|c|c}
\caption{Sequence information for $B_{22}$ = -1000 $\sigma^3$} \\
\toprule
\textbf{Index} & \textbf{Sequence} & \textbf{$x_{\text{H}}$} & \textbf{$\epsilon$} & \textbf{PS} \\
\midrule
\endfirsthead

\toprule
\textbf{Index} & \textbf{Sequence} & \textbf{$x_{\text{H}}$} & \textbf{$\epsilon$} & \textbf{PS} \\
\midrule
\endhead

\midrule
\multicolumn{5}{r}{\textit{Continued on next page}} \\
\midrule
\endfoot

\bottomrule
\endlastfoot
1 & PHPHPPHPPHPPHPPHPHPH & 0.4 & 2.442 & X \\
2 & HHPPPPHPPHHPHPHPHPPP & 0.4 & 1.871 & X \\
3 & HPPHPPHPPHPPHPPHPHPH & 0.4 & 2.385 & X \\
4 & PHPHPPHPPHPPHPPHPPHH & 0.4 & 2.166 & X \\
5 & HPHPHPPHPHPPPPHPHPPH & 0.4 & 2.296 & X \\
6 & PHPPPPPHHHHPHHPPPHPP & 0.4 & 1.506 & X \\
7 & HPPPHHHPHPPHPPPHPPPH & 0.4 & 1.830 & X \\
8 & PPHHHPHPHPPPPPHPHPPH & 0.4 & 1.840 & X \\
9 & PPHPPPHPPHPHHHHHPPPP & 0.4 & 1.438 & X \\
10 & PHHHPPHPPPHPHPPPPHPH & 0.4 & 1.920 & X \\
11 & HPHPHPHPHPHPHPHPHPHP & 0.5 & 1.960 & \checkmark \\
12 & HPPHPPHHPHPHHPHPHPHP & 0.5 & 1.687 & X \\
13 & PPPHHPHHHPHPPHPPHPHH & 0.5 & 1.407 & X \\
14 & PPHHPPHPPHPHHPHPPHHH & 0.5 & 1.445 & X \\
15 & PHPPPHHHPPHHHPHHHPPP & 0.5 & 1.270 & X \\
16 & PPHHHPPHPPHPPHPPHHHH & 0.5 & 1.340 & X \\
17 & HHHHPPHHPPPHPPHHPPHP & 0.5 & 1.335 & X \\
18 & HHHHPPPHHHHPPPPPHPPH & 0.5 & 1.170 & X \\
19 & HPHHPHHHHPHPHPPPPPHP & 0.5 & 1.260 & X \\
20 & HHHPPPHHHPPHHHPPPHHH & 0.6 & 1.085 & \checkmark \\
21 & HPHPPPHHHHPHHHPHPHHP & 0.6 & 1.125 & X \\
22 & PHPHPPHHPHPHHHHPPHHH & 0.6 & 1.112 & X \\
23 & HHHHPPHHPHHHPHPPHPHP & 0.6 & 1.100 & X \\
24 & HHHHPPPHHPHHPHHHHPPP & 0.6 & 1.040 & X \\
25 & HHHHPPHPHHHPPHHHHPPP & 0.6 & 1.037 & X \\
26 & HHHHPPHHHPPHHHHPPPPH & 0.6 & 1.022 & X \\
27 & PHHPHPHHHHHHHPPPPHPH & 0.6 & 1.010 & X \\
28 & HHHPPHHHHHPHHPPHPPHP & 0.6 & 1.030 & X \\
29 & HPPPPHHPHHHHHHHPPHPH & 0.6 & 0.990 & X \\
30 & PHPPPPHHHHHPHHPHPHHH & 0.6 & 1.020 & X \\
31 & HHHHHHPPHHHPPHPPHPHP & 0.6 & 1.010 & X \\
32 & HHHHPHHPHHPHPHHPHHHH & 0.75 & 0.860 & \checkmark \\
33 & HPHPHHHHHHHHHPHPHHHP & 0.75 & 0.793 & \checkmark \\
34 & HHPHHPHHHHHHHHHPHHPP & 0.75 & 0.769 & X \\
35 & PHPHHHHHHHHHHHHPHPPH & 0.75 & 0.735 & X \\
36 & HHHHHHHHHHPHHHPHPHPP & 0.75 & 0.727 & X \\
37 & PPHPHPHHPHHHHHHHHHHH & 0.75 & 0.720 & X \\
38 & PPHPPHHHHPHHHHHHHHHH & 0.75 & 0.708 & X \\
39 & PPHPHPHPHHHHHHHHHHHH & 0.75 & 0.703 & X \\
40 & HHHHHHHHPPPPHHHHHHHH & 0.8 & 0.674 & \checkmark \\
41 & PPHHPHHHHHHHHHHHHHPH & 0.8 & 0.676 & X \\
42 & PPHHHHHHHHHHHHHHHHPP & 0.8 & 0.644 & X \\
43 & PPPHHPHHHHHHHHHHHHHH & 0.8 & 0.640 & X \\
44 & PPPPHHHHHHHHHHHHHHHH & 0.8 & 0.614 & X \\
45 & HHHHHHHHHHHHHHHHHHHH & 1.0 & 0.497 & \checkmark \\
46 & PHPPPHHHHHPHPHPHHPHH & 0.6 & 1.097 & X \\
47 & HHPHHHPHHHPPHPPHPHPH & 0.6 & 1.143 & X \\
48 & HHPHHHHPHPPHHPHPPPHH & 0.6 & 1.120 & \checkmark \\
49 & HPHHPHHHPPPHHPHPHPHH & 0.6 & 1.227 & \checkmark \\
50 & HHHHPPPHPHHPHHPHPHPH & 0.6 & 1.168 & \checkmark \\
51 & HHPHPHHHPHPHPPPHHPHH & 0.6 & 1.197 & \checkmark \\
52 & HHHHHPPHPPPHHHPPPHHH & 0.6 & 1.028 & \checkmark \\
53 & PHPHHHHPHHPHPPPHHHHP & 0.6 & 1.117 & X \\
54 & HHPHHPHPPHPHHHPPHPHH & 0.6 & 1.218 & \checkmark \\
55 & HPHHHPHPHPHPPHHPHHHP & 0.6 & 1.247 & \checkmark \\
56 & HPHHPHHPPHHPPHHPHHPH & 0.6 & 1.303 & \checkmark \\
57 & HPHHPHPHPHHPHPHPHHPH & 0.6 & 1.368 & \checkmark \\
58 & HPHHPHHPHPHPHHPHHPHP & 0.6 & 1.326 & \checkmark \\
59 & HHHPHHPHHHPHHPHHPHHH & 0.75 & 0.880 & \checkmark \\
60 & PHHHPHHHPHHHPHHHPHHH & 0.75 & 0.865 & \checkmark \\
61 & HHHPHPHHHPPHHPHHHHHH & 0.75 & 0.818 & \checkmark \\
62 & HHPHHHHHHPHPHHPHHHHP & 0.75 & 0.835 & \checkmark \\
63 & HHHHHPHPHHPHHHPHHHPH & 0.75 & 0.850 & \checkmark \\
64 & HHHPHPHHHHHHPHHHPHHP & 0.75 & 0.822 & \checkmark \\
65 & HPHPHHHHPHHHHHPHHHHP & 0.75 & 0.822 & \checkmark \\
66 & HPHHHPHPHHHPHHHPHHHH & 0.75 & 0.857 & \checkmark \\
67 & HHPHPHHHHPHHHHHHHHHP & 0.8 & 0.729 & \checkmark \\
68 & HHPHHHHHHPHHHHHPHHPH & 0.8 & 0.747 & \checkmark \\
69 & HHHHHPHHHPHHHHPHPHHH & 0.8 & 0.758 & \checkmark \\
70 & HPHHPPHHHHHHHHHHPHHH & 0.8 & 0.710 & \checkmark \\
71 & HPPHHHHHPHHHHHHHPHHH & 0.8 & 0.720 & \checkmark \\
72 & HHHPHHHPHHHHPHHHPHHH & 0.8 & 0.776 & \checkmark \\
\end{longtable}

\subsection*{Curated HPS IDP model challenge dataset}
The following table provides the sequence information and $B_{22}$ values for HPS IDP heteropolymers that constitute the ``challenge set'' shown in \figref{fig:6}B.
The complete dataset can be found in Ref.~\cite{yaxin-wes-data-set}.

\begin{longtable}{r|c|c|c}
\caption{Sequence information for the curated HPS IDP model challenge dataset} \\
\toprule
\textbf{Index} & \textbf{Sequence} & \textbf{$B_{22}\:(\AA^3)$} & \textbf{PS} \\
\midrule
\endfirsthead

\toprule
\textbf{Index} & \textbf{Sequence} & \textbf{$B_{22}\:(\AA^3)$} & \textbf{PS} \\
\midrule
\endhead

\midrule
\multicolumn{4}{r}{\textit{Continued on next page}} \\
\midrule
\endfoot

\bottomrule
\endlastfoot
1 & RRRKKKFKWQTLWAEETEEEE & -93000 & \checkmark \\
2 & KRTRLYKLIIIFLWLLDDDDD & -92700 & \checkmark \\
3 & DTDDYYDTYTYYYFLLRKKKK & -92500 & \checkmark \\
4 & EDDDWIMFWMTHGWTKRKRRR & -92300 & X \\
5 & EEEEFYIYETYWHYFYFKRRK & -92300 & X \\
6 & RRRKMTFWWWTFLVEEFMEEE & -91700 & X \\
7 & RRKYLLPWYIILYLAYWDDDD & -91600 & \checkmark \\
8 & RWWCYWRPWYWWWRPWYPWWY & -91500 & \checkmark \\
9 & KRKKKFWQWFQWMMEIWFEEE & -91400 & \checkmark \\
10 & KRKKKTLNQWFLMLWDIDDEE & -90700 & \checkmark \\
11 & YDDDMEWTYFYMWTTKRKKYK & -90500 & \checkmark \\
12 & WDDDDILLLNVLYWPKRKKKL & -90200 & \checkmark \\
13 & PWLYWWWPFPWPPIFPWPPPP & -90000 & \checkmark \\
14 & LLPWFTPPPWWWLWWPPWGPP & -88800 & \checkmark \\
15 & RRRKYMTKWQTEWWFEFEEEE & -88700 & X \\
16 & RRRKLKFWWTTLWEEQIEETE & -88500 & X \\
17 & RRKRTMLLAPLWPYWDSDDDD & -87800 & \checkmark \\
18 & KRRKTWMYFYWWQYYFYWMEE & -87000 & X \\
19 & LDDDDWGLLCYLYFWCTRKKK & -86000 & \checkmark \\
20 & DDDMDTTLTFTCYYFYRKRRK & -85900 & \checkmark \\
21 & RKKKRLLLLKNLHLDLDDDDL & -85400 & X \\
22 & DDDLLLHWPLWLLTFWKKKRR & -85000 & X \\
23 & RRRRRILYYWRLYIYDLWPDD & -85000 & X \\
24 & KKKYKWYWKWYIGWFIWPDDD & -84700 & X \\
25 & TWEEEEYDWYTQWMNKKKKYK & -84700 & \checkmark \\
26 & KRRRYLPWYILWLRLTWDWDD & -84700 & \checkmark \\
27 & RRRKMTFWWWTWEIEETYWEE & -84600 & X \\
28 & DDDDDYGLYVILPIFGRKRKK & -84500 & \checkmark \\
29 & DDDVDAVFYLWLGWGKRRKKR & -84000 & X \\
30 & KKKRYLLIQWLILQWIEQDED & -84000 & \checkmark \\
31 & DDDIYYPPPYPPYYFRRFKKR & -83700 & X \\
32 & RKKRLGYLFPFWIVVDGDDDD & -83600 & \checkmark \\
33 & DDLDLDWWGGTPWYPLKKRKR & -83500 & \checkmark \\
34 & DLHYDDWFYLLLWRFRRFRRN & -83400 & X \\
35 & DEDMDMTLTYTWWFLITRKKK & -83200 & \checkmark \\
36 & EEEELEQEQQWHELRWRRKKR & -81800 & X \\
37 & DEFEMLFFFFFWFFFFMWKKK & -81800 & \checkmark \\
38 & EEDDLMNLMLMYYMTKRRKKL & -81600 & X \\
39 & DDDLYLWFYWWPLPPWPRKPK & -81300 & \checkmark \\
40 & DDDDLALWWLGPLRIGPRRRK & -80900 & \checkmark \\
41 & DDEMDTTLTFTWWFLIVRKKK & -80700 & \checkmark \\
42 & KKTKRYMGLYWHPWDDDYDDD & -80700 & X \\
43 & EQEEEFFMFWQQFQWKKWKKR & -79900 & X \\
44 & YDDDDMHLTFRWMFLIPRKKK & -79600 & \checkmark \\
45 & RRIRRIGCWWPWWPLWYVWDD & -79500 & X \\
46 & RMRKKKFKWMTKWTEETEEEE & -79400 & X \\
47 & DDDMQAFLTFTWWYLIMRKKK & -79100 & X \\
48 & RRRKTTTWYWTFWPEEFVDEF & -79000 & \checkmark \\
49 & KMRKRFYWQHFYEWEQEEEWQ & -78500 & X \\
50 & EEEEWIHYTTYWYYFYIKPRK & -78300 & X \\
51 & RRRRRTTYYTYLPTAPPDDDD & -77900 & \checkmark \\
52 & DDDDDLWFIFMPWMYTTTRRK & -77900 & X \\
53 & RRRRKKFKGCTLWAEEEEGEE & -77800 & X \\
54 & DDDDDYWYPYPWPPFYKGRIK & -77700 & X \\
55 & RTRFRRFKTQTYWIEETEEEE & -77600 & X \\
56 & DDDTDTTLVYYYTTLLYKRKR & -77600 & \checkmark \\
57 & RRRKKFFMHNHLTAEEMEEEW & -77400 & X \\
58 & IRRKKNFKWQTLWAEETEEEE & -77400 & X \\
59 & RKKKIPPWIPGYWADYWGDDD & -77300 & \checkmark \\
60 & DEDDDSIPLTLGTYHWTKRRR & -77200 & X \\
61 & MRRKKKFKWQTLWIEFTEEEE & -77100 & X \\
62 & RRRKKRFMWTTTAFPETEEEE & -77000 & X \\
63 & DDDYLYYLDLYQLLLLRLRRR & -77000 & \checkmark \\
64 & KKKYKSWLLLLYLLILWDDWD & -76100 & \checkmark \\
65 & KKRKKMFKWQTLGAEETEEEE & -76000 & X \\
66 & WRKKNGWTTVLYYYYDWDDDD & -75200 & X \\
67 & KRRYRINLYYISMVDDGDLDD & -75000 & X \\
68 & KKNKKKLLLLLIYILEDDYDY & -74800 & X \\
69 & DDDDDFLNNINNWLNRNRRKN & -74400 & X \\
70 & DWDDGLLIPIWIIWIRIKWRR & -73800 & \checkmark \\
71 & DDDPDYPIYIGMIPYIRRCRR & -73200 & \checkmark \\
72 & DDDIDIILAPVWWIIWLPRKK & -73200 & X \\
73 & TWDDPWPYTWPYYWWPWPKKK & -72200 & \checkmark \\
74 & LWDDHWWLLIYLLLFLLRRKW & -72100 & \checkmark \\
75 & RRIKRYLLFNLLILTNVEDDD & -71400 & \checkmark \\
\end{longtable}

\subsection*{Sequences ranked by variance divergence index}

The sequences below are ranked by their variance divergence index, \eqref{eq:vdi}.
Larger values indicate that the contact-map variance is more spread across high-wavenumber modes.
Table~\ref{tab:top10} lists the top 10\% of sequences by variance divergence index in the combined $B_{22}$ dataset.
Table~\ref{tab:bottom10} lists the bottom 10\%, which serves as a control subset in our analysis.
Each sequence’s original table and index are noted for reference.
The complete dataset can be obtained at \texttt{https://github.com/wmjac/} \texttt{heteropolymer-phase-separation}.

\begin{longtable}{c|c|c}
\caption{Sequences with highest variance divergence index} \\
\toprule
\textbf{Sequence} & \textbf{$B_{22}\:(\AA^3)$} ($\sigma^3$) & \textbf{Table} \\
\midrule
\endfirsthead

\toprule
\textbf{Sequence} & \textbf{$B_{22}\:(\AA^3)$} ($\sigma^3$) & \textbf{Table} \\
\midrule
\endhead

\midrule
\multicolumn{3}{r}{\textit{Continued on next page}} \\
\midrule
\endfoot

\bottomrule
\endlastfoot
HPHPHPHPHPHPHPHPHPHP & -1000 & T2 (Row 11) \\
HHHHPPHPHHHPPHHHHPPP & -1000 & T2 (Row 25) \\
HHHHPPHPHHHPPHHHHPPP & -400 & T1 (Row 26) \\
HHHPPPHHHPPHHHPPPHHH & -1000 & T2 (Row 20) \\
HHHHPPPHHPHHPHHHHPPP & -1000 & T2 (Row 24) \\
HHHHPPPHPHHPHHPHPHPH & -1000 & T2 (Row 50) \\
HPHHPHHPPHHPPHHPHHPH & -400 & T1 (Row 60) \\
HPHHPHPHPHHPHPHPHHPH & -400 & T1 (Row 61) \\
PHPHPPHPPHPPHPPHPHPH & -1000 & T2 (Row 1) \\
PHPHPPHPPHPPHPPHPPHH & -1000 & T2 (Row 4) \\
HHHPHHPHHHPHHPHHPHHH & -1000 & T2 (Row 59) \\
HHHHPPPHPHHPHHPHPHPH & -400 & T1 (Row 52) \\
HHPHHPHPPHPHHHPPHPHH & -1000 & T2 (Row 54) \\
HHPHHPHPPHPHHHPPHPHH & -400 & T1 (Row 58) \\
HHHHPPPHHPHHPHHHHPPP & -400 & T1 (Row 25)
\label{tab:top10}
\end{longtable}

\newpage
~\\

\begin{longtable}{c|c|c}
\caption{Sequences with lowest variance divergence index} \\
\toprule
\textbf{Sequence} & \textbf{$B_{22}\:(\AA^3)$} ($\sigma^3$) & \textbf{Table} \\
\midrule
\endfirsthead

\toprule
\textbf{Sequence} & \textbf{$B_{22}\:(\AA^3)$} ($\sigma^3$) & \textbf{Table} \\
\midrule
\endhead

\midrule
\multicolumn{3}{r}{\textit{Continued on next page}} \\
\midrule
\endfoot

\bottomrule
\endlastfoot
PPHPPPHPPHPHHHHHPPPP & -400 & T1 (Row 9) \\
PHPHHHHHHHHHHHHPHPPH & -400 & T1 (Row 36) \\
PHPHHHHHHHHHHHHPHPPH & -1000 & T2 (Row 35) \\
PPHPPPHPPHPHHHHHPPPP & -1000 & T2 (Row 9) \\
PHPPPPPHHHHPHHPPPHPP & -1000 & T2 (Row 6) \\
PHHPHPHHHHHHHPPPPHPH & -400 & T1 (Row 28) \\
PHPPPPPHHHHPHHPPPHPP & -400 & T1 (Row 6) \\
HPHPHHHHHHHHHPHPHHHP & -400 & T1 (Row 34) \\
PPHHPHHHHHHHHHHHHHPH & -400 & T1 (Row 42) \\
PPHHHHHHHHHHHHHHHHPP & -400 & T1 (Row 43) \\
PPHHHPHPHPPPPPHPHPPH & -400 & T1 (Row 8) \\
HPPPPHHPHHHHHHHPPHPH & -400 & T1 (Row 30) \\
HPHHPPHHHHHHHHHHPHHH & -400 & T1 (Row 73) \\
PHHPHPHHHHHHHPPPPHPH & -1000 & T2 (Row 27) \\
PPHHPHHHHHHHHHHHHHPH & -1000 & T2 (Row 41)
\label{tab:bottom10}
\end{longtable}

\end{document}